\documentclass[mathleft
]{an}
\usepackage{graphicx}
\usepackage{times}
 \usepackage[sort]{natbib}
 \bibpunct{(}{)}{;}{a}{}{,}
\usepackage{subfigure}
\usepackage{amsmath}
\usepackage{color}
\definecolor{darkgreen}{RGB}{0,142,128}

\begin{document}
\Pagespan{1}{}
\Yearpublication{2011}%
\Yearsubmission{2012}%
\Month{1}%
\Volume{1}%
\Issue{1}%

\title{Magnetic confinement of the solar tachocline:\\ The oblique
  dipole.}

\author{A. Strugarek \inst{1}
 \fnmsep
  \thanks{Corresponding author:
  \email{antoine.strugarek@cea.fr}\newline}, A. S. Brun \inst{1}
\and  J.-P. Zahn\inst{2,1}
}
\titlerunning{Magnetic confinement of the solar tachocline}
\authorrunning{A. Strugarek \textit{et al.}}
\institute{
Laboratoire AIM Paris-Saclay, CEA/Irfu Université Paris-Diderot CNRS/INSU, 91191 Gif-sur-Yvette, France
\and 
LUTH, Observatoire de Paris, CNRS-Université Paris Diderot, Place
Jules Janssen, 92195 Meudon, France}

\received{30 May 2011}
\accepted{11 Nov 2011}
\publonline{later}


\abstract{%
 3D MHD global solar simulations coupling the turbulent convective zone and the radiative zone have been carried out. Essential features of the Sun such as differential rotation, meridional circulation and internal waves excitation are recovered. 
These realistic models are used to test the possibility of having the
solar tachocline confined by a primordial inner magnetic
field. We find that the initially confined magnetic fields we consider open
into the convective envelope. Angular momentum is transported across
the two zones by magnetic torques and stresses, establishing the so-called
Ferarro's law of isorotation. In the parameter space studied, the confinement of the magnetic field by
meridional circulation penetration fails, also implying the failure of
the tachocline confinement by the magnetic field. Three-dimensional
convective motions are proven responsible for the lack of magnetic
field confinement. Those results are robust for the different
magnetic field topologies considered, i.e. aligned or oblique dipole.}

\maketitle

\section{Introduction}
\label{sec:introduction}

Since its discovery \citep[][]{Brown:1989hq}, the solar tachocline
has puzzled the scientific community. In particular, a clear
explanation for its
extreme thinness \citep[less than 5\% of the solar radius $R_\odot$, see][]{Charbonneau:1999es} is still lacking.
The first theoretical work dedicated to the tachocline was carried out by
\citet{Spiegel:1992tr}. They considered an hydrodynamic tachocline and
showed that such an interface layer will spread into the radiative
interior because of thermal diffusion. As a result, they estimated
that the differential rotation should extend down to $0.3\, R_\odot$
after $4.5\, $Gyears, which is in total contradiction with the
helioseismic inversions
\citep{Schou:1998bi,Thompson:2003bk}. Consequently, additional
physical mechanisms have to be considered in order to properly explain
the observed thinness of the tachocline.

\citet{Spiegel:1992tr} suggested that the stratification in the
tachocline would imply an anisotropy in
the turbulence, making it predominantly horizontal. It would then erode the
latitudinal gradient of angular-velocity \citep{Elliott:1997wz}. However
\citet{Gough:1998ik} pointed out that such anisotropic turbulence would
on the contrary act as an anti-diffusion \citep[][]{Dritschel:2008em}. 
In fact, the question is still hardly settled when both
radial and latitudinal shears of angular-velocity are taken into
account \citep{Miesch:2003eh,Kim:2005fe,Leprovost:2006ft,Kim:2007cy}.
Further, \citet{Tobias:2007ks} showed that the introduction of (even weak)
magnetic fields in the bulk of the tachocline would erase both the
diffusive and anti-diffusive behaviors of 2D turbulence. 

\citet{Gough:1998ik} then proposed that a (fossil) dipolar magnetic field confined
in the solar interior could oppose the thermal spreading of the
tachocline. This solution offered also an explanation
for the solid body rotation of the solar radiation zone and such a fossil field was also invoked by
\citet{Rudiger:1997wz} to confine the tachocline. 
The fossil field confinement scenario is in fact a double
confinement problem. First, the imposed magnetic field erodes
latitudinal gradients of angular-velocity, thus confining the tachocline. Second,
the magnetic field has to remain confined in the radiation zone
against its outward ohmic diffusion.
In order to confine the magnetic field, it was first argued that a meridional flow coming
from the convection zone down to the radiative interior through the
tachocline at the high latitudes could eventually prevent the field from diffusing outward in this
region \citep{Gough:1998ik}. At the equator,  \citet{Wood:2011es} suggested that if the
magnetic field were to connect to the convection zone, the magnetic
pumping \citep{Tobias:2001ho,Dorch:2001jx,Ziegler:2003fd} would
'confine' it below the
region of intense shear.

Despite the appeal of this scenario, none of the many numerical
simulations that have been carried out in the past
succeeded in recovering it completely. The first simulations of the
radiation zone in 2D \citep{Garaud:2002fq} and in 3D
\citet{Brun:2006bb} showed that if the
confinement of the magnetic field fails, angular momentum is
transported along the field lines into the radiation zone, making the
radiative interior rotate differentially. The Ferraro's
law of iso-rotation is then established \citep{Ferraro:1937ts}, in complete
contradiction with solar observations.
However these models did not allow for flows penetrating from the
convection zone, such as plumes or meridional circulation.
In their simulations, \citet{Sule:2005bq,Rudiger:2007be,Garaud:2008fi}
imposed a meridional circulation at the
top of the radiation zone. They recovered partially the \citet{Gough:1998ik}
scenario, but one may object that those results are highly sensitive to
the prescribed (profiles and amplitude) meridional circulation. Simulations
coupling self-consistently the two zones of the Sun were then carried
out by \citet{Rogers:2011kw} in 2D and by
\citet{Strugarek:2011cx} (hereafter SBZ11) in 3D in order to take into
account the motions of the convection zone. Both studies were not able
to validate the
magnetic confinement scenario. Finally, \citet{Wood:2011id} conducted
an analytical study in a reduced cartesian model and proposed an
improved theory for a magnetic confinement scenario \textit{\`a la} \citet{Gough:1998ik}.

We are interested here in the fact that SBZ11 showed
that the convective motions were responsible for the
lack of confinement of the buried magnetic field, and that in most of
the works previously cited, the magnetic field
enters the convection zone primarily at the equator. Since different
magnetic topologies may lead to significantly different interactions
with the convective motions, the universality of the results obtained
with our axisymmetric dipolar field can be questioned.
We thus conduct here numerical simulations based on the SBZ11 model to investigate the role of
the magnetic field topology. In Sect. \ref{sec:model-both-conv} we
summarize the main ingredients of the model used by
SBZ11, and in Sect. \ref{sec:infl-magn-topol} we
study the impact of the magnetic topology on the confinement of the
field. Conclusions and perspectives are reported in Sect. \ref{sec:consl-persp}.

\section{Modeling the convection and the radiation zones}
\label{sec:model-both-conv}

Following \citet{Brun:2011ji}, SBZ11, we use the ASH code
\citep{Brun:2004ji} to model
$90\%$ of the solar interior in 3D using the non-linear MHD equations under the
anelastic approximation. A LES (large eddy simulation) approach is
used to parametrize turbulent diffusivity profiles; the detailed
equations can be found in SBZ11. 

\begin{figure}
  \centering
  \includegraphics[width=.65\linewidth,angle=90]{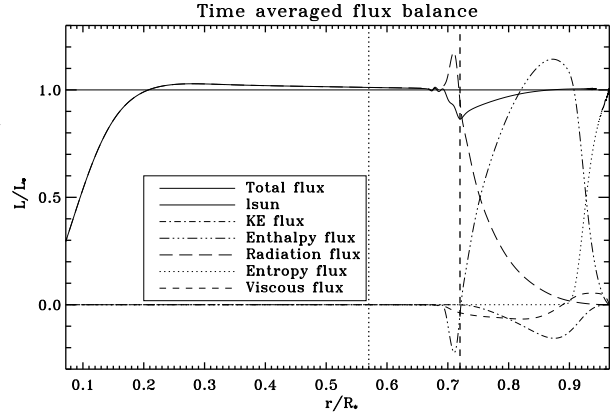}
 \caption{Radial energy flux balance normalized to the luminosity of
   the Sun. The dashed vertical line at $r=r_{bcz}/R_\odot$ denotes the
   base of the convection zone, the dotted vertical line at
   $r=R_b/R_\odot$ denotes the bounding radius of the initial magnetic
 field (see Eq. \eqref{eq:Psi_Choice}).}
  \label{fig:flux_bal}
\end{figure}

Our initial hydrodynamical setup is exactly the same than the one used in SBZ11. We
recall here the principal features of the model. We use the solar
rotation rate $\Omega_0 = 2.6\,10^{-6} s^{-1}$, the solar mass
$M_\odot=2.0\, 10^{33}\, g$ and the solar luminosity $L_\odot=3.8\, 10^{33}\,
erg\, s^{-1}$. The background
thermodynamic quantities were computed with the CESAM code
\citep{Morel:1997gy}, hence we use the solar stratification and obtain
a realistic Brunt-V\"ais\"al\"a frequency $N$ in the radiation zone. The radial gradient of
entropy is negative above $r_{bcz}=0.715\, R_\odot$, it defines the
convection zone since the Rayleigh number ($\sim 10^5$) in this region is
well above the critical Rayleigh number for the onset of the
convective instability \citep{Gilman:1981eh}. We display in Fig. \ref{fig:flux_bal} the
radial energy flux balance normalized to $L_\odot$. The energy is carried by the
radiative flux in the radiation zone (below $r_{bcz}$, the black
dashed line), and the enthalpy flux transports the major part of the energy
in the convection zone (above $r_{bcz}$). 

The convective motions exhibit
banana-like shapes at the equator \citep{Brun:2002gi}, and more patchy
patterns at higher latitudes. Angular momentum is transported mainly by
convective Reynolds stresses in the bulk of the convection zone,
establishing a differential rotation similar to the solar differential
rotation (see for example the color contours in
Fig. \ref{fig:polar_dip}). A tachocline develops, and the angular
velocity shear extends down to $r=0.58\, R_\odot$. The meridional circulation which is
self-consistently excited in the convection zone is roughly
unicellular in each hemisphere (when time-averaged), with downflows
near the poles and upflows at the equator. We stress here that those
mean flows are self-consistently induced by the interaction between
the convective motions, the rotation, and the baroclinicity induced by
latitudinal gradients of entropy. Thus, they are not artificially imposed
by parameters, by boundary conditions or prescribed profiles. The convective overshooting
depth is $d_{ov}=0.04 \, R_\odot$ (see the enthalpy flux in
Fig. \ref{fig:flux_bal}) and is known to scale as the square root of the filling
factor of downflow plumes \citep{Zahn:1991uz}. Consequently, it is certainly
overestimated in this model. The mean meridional circulation penetrates by
$d_{MC}=0.035\, R_\odot$ below the base of the convection zone
$r_{bcz}$. Even if our model is not completely in the solar parameter
regime due to our enhanced diffusivities (see SBZ11), we observe a
penetration of both the meridional circulation and the convection below $r_{bcz}$. We are
thus confident that our model captures the key physical ingredients
of the upper tachocline. Finally, we observe also that the convective overshooting plumes
excite gravity waves in the radiation zone \citep{Brun:2011ji}.

Since we used enhanced viscosity $\nu$, thermal diffusivity $\kappa$
and magnetic diffusivity $\eta$, the time-scales involved in the
simulation differ from what is occurring in the real Sun. Still we
took great care to maintain the proper hierarchy between the time-scales,
though they are somewhat closer to each other than what they are in
reality. The ohmic time-scale $\tau_\eta=R_\odot^2/\eta$ is
$\tau_{\eta}^{CZ}=9.60$ years in the convection zone, and
$\tau_{\eta}^{RZ}=192$ years in the radiation zone. By comparison, the
convective turnover time is $\tau_{conv} = 28$ days. We refer the
reader to SBZ11 for further details on the simulations.

\section{Influence of the magnetic topology}
\label{sec:infl-magn-topol}

\subsection{On the axisymmetric dipole topology}
\label{sec:opening-field-at}

The expression of the magnetic field used in SBZ11
is given by $\mathbf{B} = B_0\left(B_r\mathbf{e}_r+
  B_\theta\mathbf{e}_\theta\right)$, with 
\begin{equation}
  \label{eq:mag_field_init}
  B_r = \frac{1}{r^2\sin\theta}\partial_\theta\Psi \, , \hspace{1cm}
  B_\theta = -\frac{1}{r\sin\theta}\partial_r\Psi \, ,
\end{equation}
where $\Psi(r,\theta) = r\sin\theta A_\varphi$ is constant on field lines and is such that 
\begin{eqnarray}
  \label{eq:Psi_Choice}
  \Psi &= \left({r}/{R}\right)^2(r-R_b)^2\sin^2\theta & \mbox{for
  } r\le R_b \, , \nonumber\\
  &= 0 & \mbox{for } r\ge R_b \, ,
\end{eqnarray}
where $R_b=0.57\, R_\odot$ is the bounding radius of the confined field.
One can immediately see that for such a choice of $\Psi$, the maximum
radial gradient of $B_\theta$ is located at the equator. Since the
behavior of the magnetic field is mainly diffusive in the bulk of the
radiation zone, it will evolve more rapidly where its gradients are
maximum. As a result, it preferentially interacts with the tachocline (and
eventually the convection zone) at the equator.

\begin{figure}
  \centering
  \includegraphics[width=.45\linewidth]{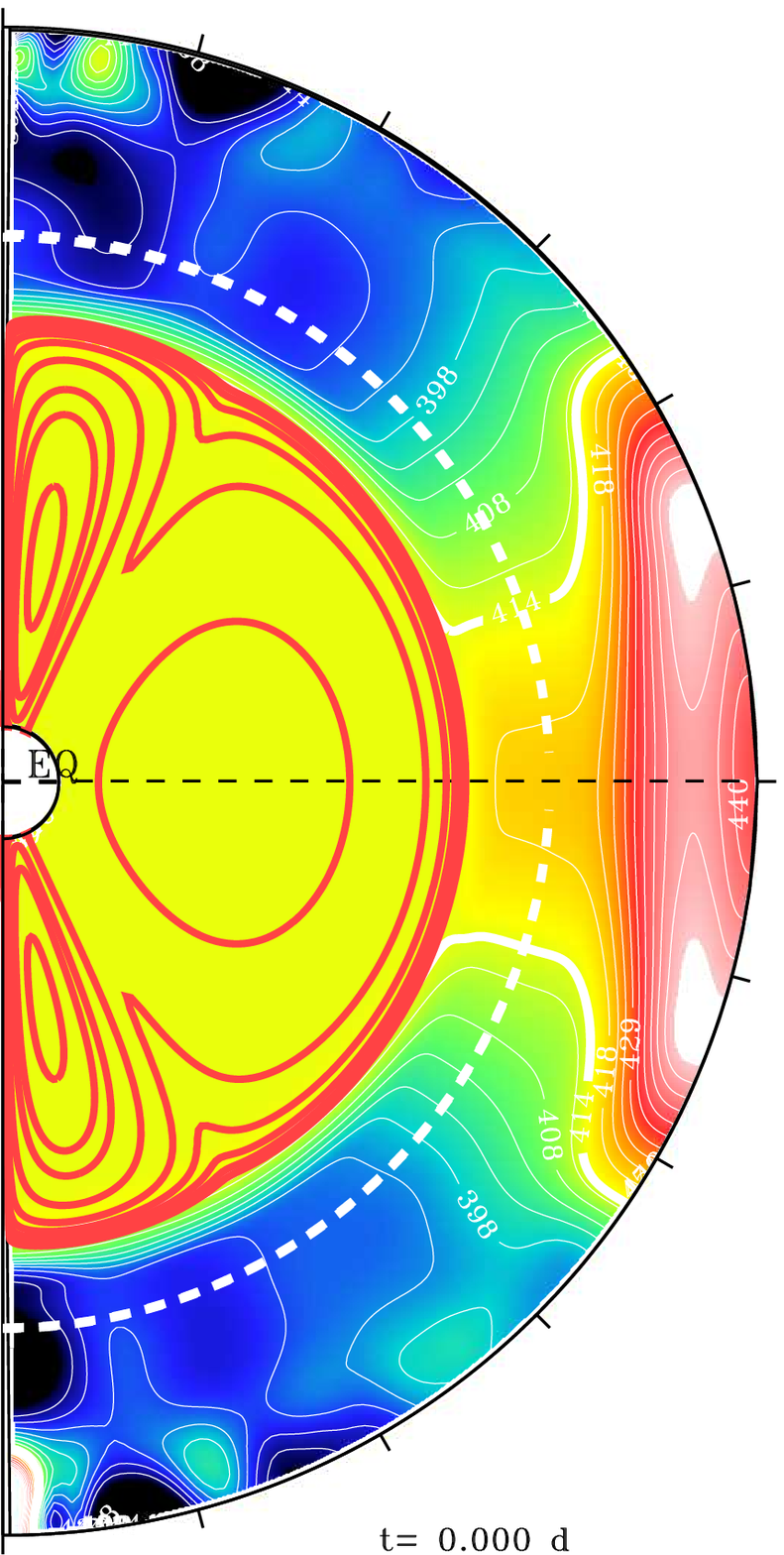}
  \includegraphics[width=.45\linewidth]{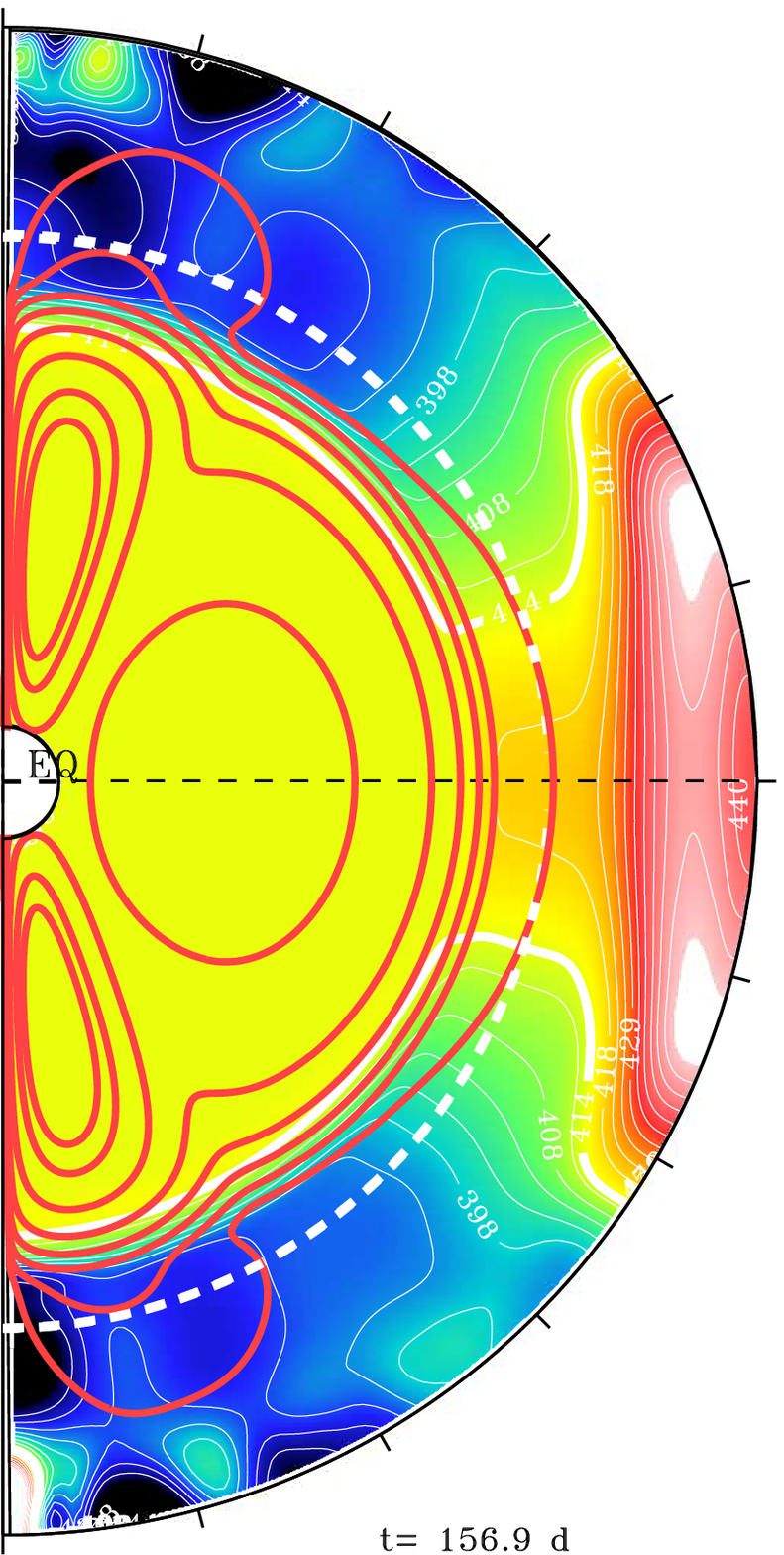}
  \caption{Azimuthal averages of the angular velocity $\Omega$ (color
    map) and the magnetic field lines (red lines), in the axisymmetric
    case. The $\Omega$
    contours have also been temporally averaged.}
  \label{fig:polar_dip}
\end{figure}

One may construct modified 'dipoles' in order to move the location of
maximum gradient with latitude. To do so, we consider a $\Psi$ function
defined by
\begin{equation}
  \label{eq:modif_grad_dipole}
  \Psi = \left[ \alpha\sin^2\theta +
    (1-\alpha)\sin^2\theta\cos^{2p}\theta \right]f(r)\, ,
\end{equation}
 where $\alpha$ and $p$ are related to the co-latitude $\theta_{max}$
 where the gradient of $B_\theta$ is maximum, and $f(r)$ controls the
 radial shape of the dipole. We choose $\theta_{max}=10^\circ$ and run the
 ASH code. We display in Fig. \ref{fig:polar_dip} the initial magnetic
 configuration, and the evolution of the magnetic field at a later instant.
We observe that the dipole interacts with the convective motions primarily at latitude
$80^\circ$, as expected. This confirms the key role played by the
field topology in controlling the temporal evolution of the field in
the stable radiative interior.

Nevertheless, the end state in this case will be the same than in
SBZ11. Even if a longer time is needed for the
magnetic field lines to interact with the convective motions at the
equator, the modified dipole is still axisymmetric. As a result, its
interaction with the convective zone will be similar to the behavior
described in SBZ11. Angular momentum will still be transported along
the field lines, although slight differences may occur due to the
little topological change in the bulk of the radiative zone. This will
result again in a differentially rotating radiative zone. In the end,
the magnetic scenario for the solar tachocline will again certainly
fail in this case. 

\subsection{Non axisymmetric magnetic topology: the oblique dipole}
\label{sec:diff-magn-topol}

The oblique magnetic fields in stars have been a longstanding subject of research
\citep{Mestel:1972tr,Mestel:1987wa}. For example, A-type stars are
typically thought to be oblique rotators \citep{Brun:2005fv}.
Given that a purely axisymmetric field establishes Ferraro's law, one may wonder whether an oblique dipole
acts similarly. In addition, any confined dipolar magnetic
field may explain the solid body rotation of the radiation zone. We
consequently test the robustness of the results of SBZ11 by
considering a tilted dipole. 

\begin{figure}
  \centering
  \includegraphics[width=\linewidth]{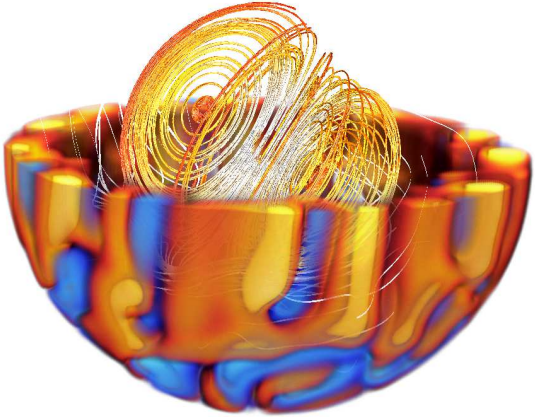}  
  \caption{3D rendering of magnetic field lines. The volume rendering
    is the azimuthal velocity in the rotating frame in the south
    hemisphere, orange denotes
    positive $v_\varphi$ and blue negative $v_\varphi$.}
  \label{fig:3D_ob_dip}
\end{figure}

We use the ASH formalism and decompose the magnetic field into
poloidal and toroidal components,
\begin{equation}
  \mathbf{B} =
  \boldsymbol{\nabla}\times\boldsymbol{\nabla}\times\left(C\mathbf{e}_r\right)
  + \boldsymbol{\nabla}\times\left(A\mathbf{e}_r\right) \, .
  \label{eq:B_decompo}
\end{equation}
The dipole aligned with the rotation axis is then simply written
\begin{equation}
  \label{eq:dipole_simple_ASH}
  \left\{
  \begin{array}{lcl}
    A &=& 0 \\
    C &=& f(r) Y_0^1
 \end{array}
  \right. \, ,
\end{equation}
where $Y_l^m$ stands for the classical spherical harmonics. To tilt the
dipole with an angle $\beta$ with respect to the rotation axis, we simply write
\begin{equation}
  \label{eq:dipole_nax_ASH}
  \left\{
  \begin{array}{lcl}
    A &=& 0 \\
    C &=& f(r) \left( \cos\beta Y_0^1 + \frac{\sin\beta}{\sqrt{2}}Y_1^1\right)
 \end{array}
  \right. \, ,
\end{equation}
where $f(r)=B_0\left(r/R_\odot\right)^2\left(r-R_b\right)^2$ and $R_b =
0.57\, R_\odot$ is the bounding radius of the initial magnetic field.
We choose $\beta=60^\circ$ to tilt significantly the dipole. A 3D
rendering of the magnetic field lines is
displayed in Fig. \ref{fig:3D_ob_dip}.

\begin{figure}[!Hbtp]
  \centering
 \subfigure[]{
  \label{fig:AZAV_i}
  \includegraphics[width=.45\linewidth]{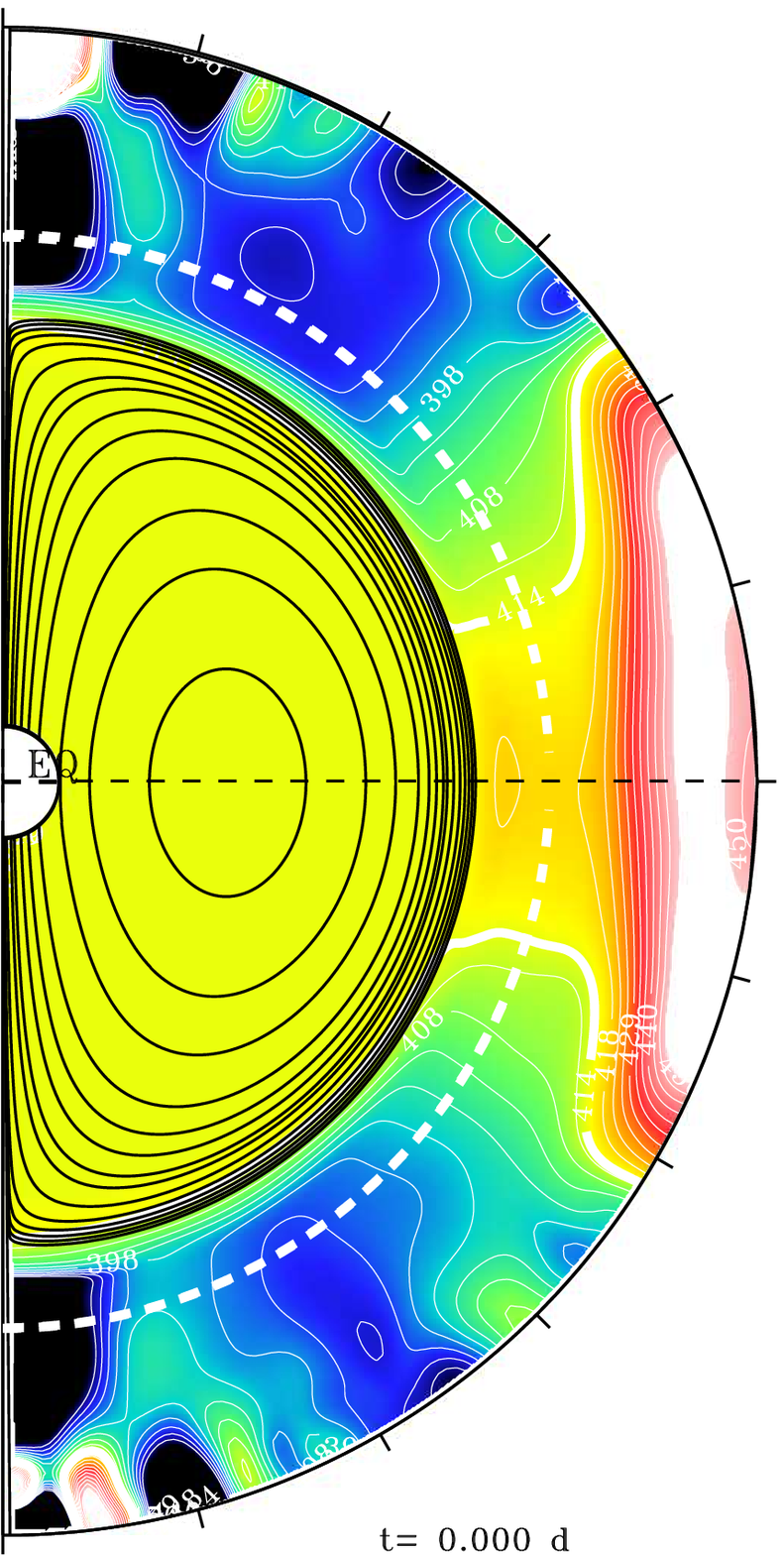}    
}
 \subfigure[]{
  \label{fig:AZAV_e}
  \includegraphics[width=.45\linewidth]{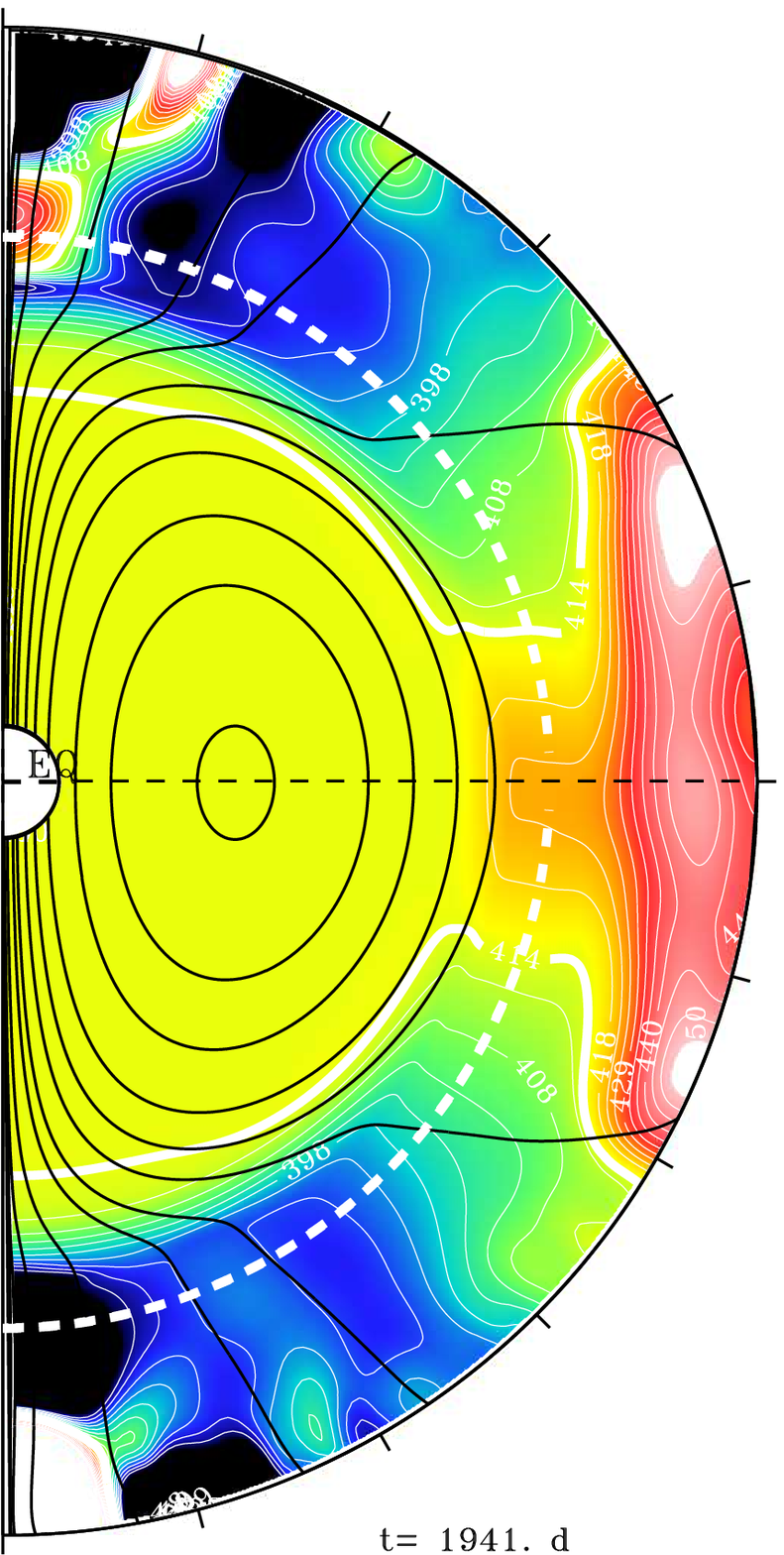}    
}
 \subfigure[]{
  \label{fig:POSL_i}
  \includegraphics[width=.45\linewidth]{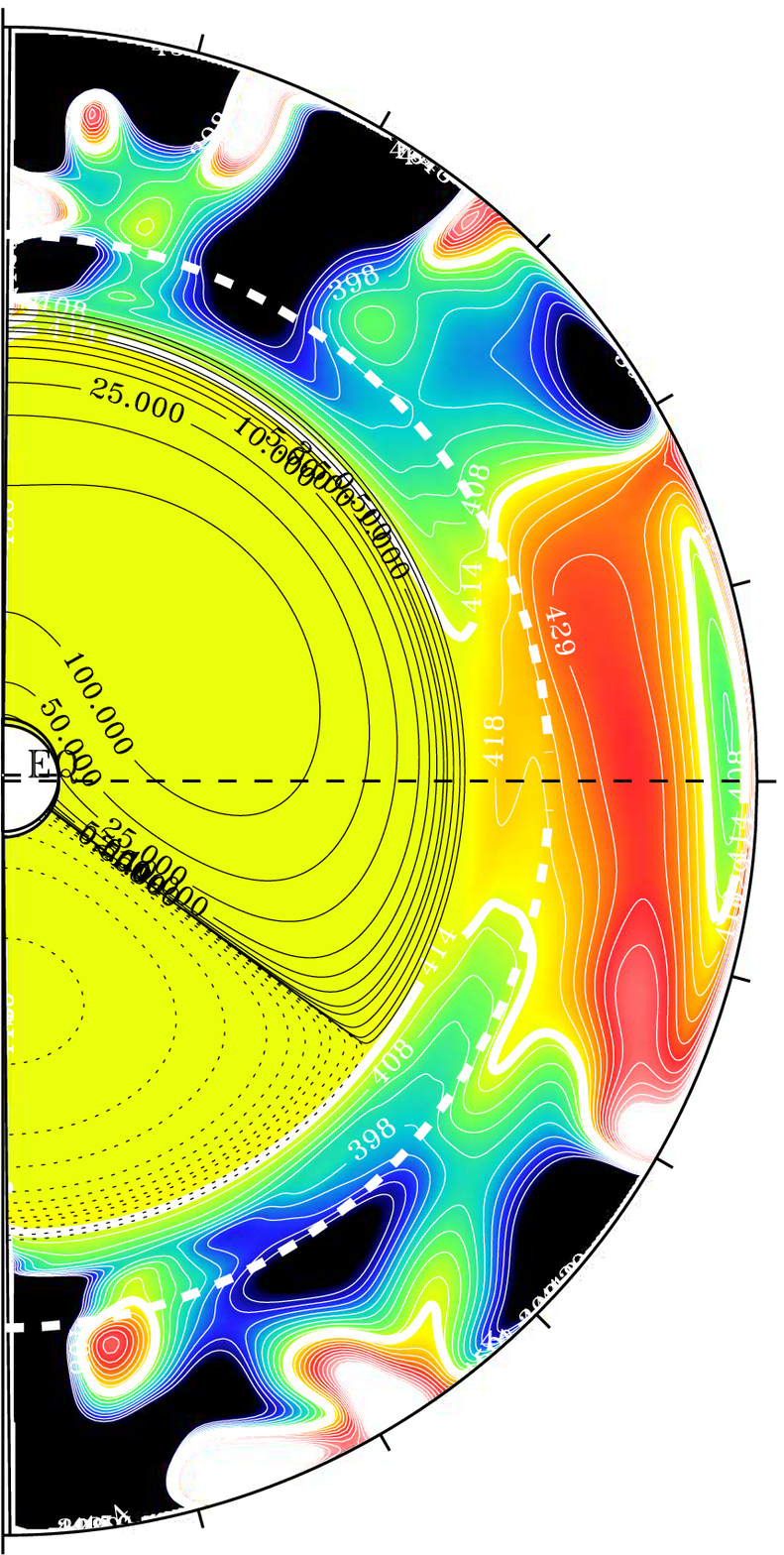}    
}
 \subfigure[]{
  \label{fig:POSL_e}
  \includegraphics[width=.45\linewidth]{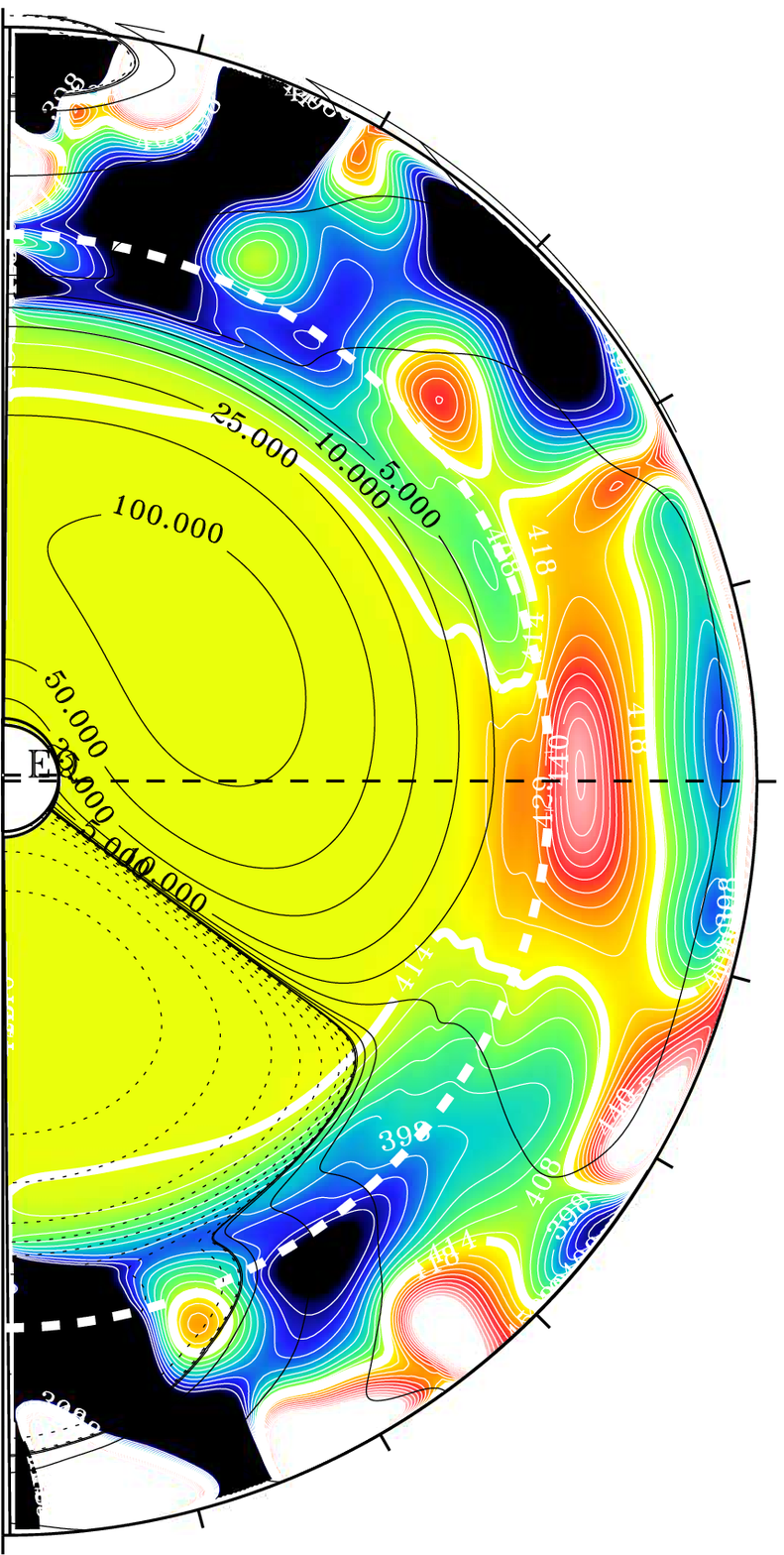}    
}
 \caption{Meridional cuts of the titled dipole case. The quantities in
   \textbf{(a,b)} are azimuthally averaged, while they are taken at
   the longitude $\varphi\sim 0$ (\textit{i.e.}, the
    longitude where the projection of the magnetic field on a polar slice
    is indeed a planar dipole, at least initially) in panels
    \textbf{(c,d)}. The slight discrepancy between the dipole angle
    and $60^\circ$ (or $-30^\circ$) is due to the fact that we are not exactly at $\varphi=0$.
     The color maps display the angular velocity $\Omega$, and the black lines the
     magnetic field lines. \textbf{(a,c}) are taken at
    the initialization of the magnetic field, and \textbf{(b,d})
    approximately $70$ convective turnover times later. }
  \label{fig:omega_fieldlines_OCdip}
\end{figure}

Although the magnetic topology is clearly different from
SBZ11, we observe that if $\beta \neq 90^\circ$, then
the initial magnetic field is the sum of an axisymmetric dipole
(\textit{i.e.}, similar to SBZ11) and a non-axisymmetric
component (\textit{see} eq. \eqref{eq:dipole_nax_ASH}). This is also made
clear by comparing Figs. \ref{fig:AZAV_i} and \ref{fig:POSL_i} which are taken
at the initialization of the magnetic field. The
azimuthally averaged field displayed in Fig. \ref{fig:AZAV_i} only retains the
axisymmetric component of the tilted dipole, although the real magnetic
configuration is oblique (Fig. \ref{fig:POSL_i}).  

Figs. \ref{fig:AZAV_e} is taken $\sim 70$ convective turnover times later.
We see that the axisymmetric component of the oblique dipole
roughly evolves similarly to the pure axisymmetric dipole of
SBZ11. We observe indeed that the initially buried magnetic field does
not remain confined and connects the convection
zone with the radiative interior. The non axisymmetric field likewise does not
remain confined in Fig. \ref{fig:POSL_e}, and its evolution is
highly three dimensional. We stress that the field lines plotted in
the radiation zone are a fair representation of the field topology,
whereas in the tachocline and in the convection zone only the
projection into the meridional plane of a 3D magnetic field
with non zero azimuthal component is represented. 

The magnetic Reynolds number realized in the convection zone is below the threshold for a
dynamo to occur ($Rm \sim 100$ \textit{vs} $300$ in \citet{Brun:2004ji}).
As a result, the total magnetic energy decreases as the simulation
evolves.

The polar slices in Fig. \ref{fig:omega_fieldlines_OCdip}
seem to exhibit major changes in the $\Omega$ profile
at the poles, and we know from SBZ11 that the unconfined axisymmetric magnetic field will apply a torque
leading to efficient transport of angular momentum. The evolution of the
non-axisymmetric component may also have some effects on the angular momentum
balance. Thus we examine the different terms of the
angular momentum balance which are
defined as follows \citep{Brun:2004ji}:
{\setlength{\mathindent}{0pt} 
\begin{equation}
  \label{eq:ang_mom_bal}
  \partial_t (\bar{\rho}\mathcal{L}) = - \mbox{\boldmath
    $\nabla$}\cdot( \underbrace{ \mathbf{F}^{MC} + \mathbf{F}^{RS} +
    \mathbf{F}^{VD} }_{\mbox{Hydro}} + \underbrace{\mathbf{F}^{MT} +
    \mathbf{F}^{MS}}_{\mbox{MHD}} ) \, ,
\end{equation}}
 where the different terms correspond respectively to contributions
 from Meridional Circulation, Reynolds Stress, Viscous Diffusion,
 Maxwell Torque and  Maxwell
 Stress. They are defined by 
{\setlength{\mathindent}{0pt} 
\begin{eqnarray}
   \label{eq:MC}
   \mathbf{F}^{MC} &\equiv&
   \bar{\rho}\langle\mathbf{v}_M\rangle\mathcal{L} \, ,\\
   \label{eq:RS}
   \mathbf{F}^{RS} &\equiv&
   r\sin\theta\bar{\rho}\left(\langle v_r'v_\varphi'\rangle\mathbf{e}_r +
     \langle v_\theta'v_\varphi'\rangle\mathbf{e}_\theta \right) \, ,\\
   \label{eq:VD}
   \mathbf{F}^{VD} &\equiv&
   - \nu\bar{\rho}r^2\sin\theta\left\{ \partial_r\left(\frac{\langle v_\varphi \rangle}{r}\right)\mathbf{e}_r +
   \partial_\theta\left(\frac{\langle v_\varphi
       \rangle}{\sin\theta}\right)\mathbf{e}_\theta\right\} \, ,\\
   \label{eq:MT}
   \mathbf{F}^{MT} &\equiv&
   -\frac{r\sin\theta}{4\pi}\langle
   B_\varphi\rangle\langle\mathbf{B}_M\rangle \, \mathrm{            and}\\
   \label{eq:MS}
   \mathbf{F}^{MS} &\equiv&
   -\frac{r\sin\theta}{4\pi}\left(\langle B_r'B_\varphi'\rangle\mathbf{e}_r +
     \langle B_\theta'B_\varphi'\rangle\mathbf{e}_\theta \right) \, ,
\end{eqnarray} }
where the subscript $._M$ designates the meridional component of $\mathbf{v}$ and $\mathbf{B}$.  In the previous equations, we
have decomposed the velocity and the magnetic field into an azimuthally
averaged part $\langle .\rangle$ and $\varphi$-dependent part (with
a prime). The different contributions can further be separated between radial
($\mathcal{F}_r$ along $\mathbf{e}_r$) and latitudinal
($\mathcal{F}_\theta$ along
 $\mathbf{e}_\theta$) contributions. However here we only focus on the
 radial flux of angular momentum defined by
\begin{eqnarray}
 \label{eq:poloidal_flux_ang_mom}  
  \mathcal{I}_r(r) &=& \int_{\theta_1}^{\theta_2}
  \mathcal{F}_r(r,\theta)r^2\sin\theta \hspace{0.2cm}\mbox{d}{\theta}
\end{eqnarray}
where $(\theta_1,\theta_2)$ maybe chosen to study a particular
region of our simulation.

\begin{figure}
  \centering
  \includegraphics[width=.8\linewidth,angle=90]{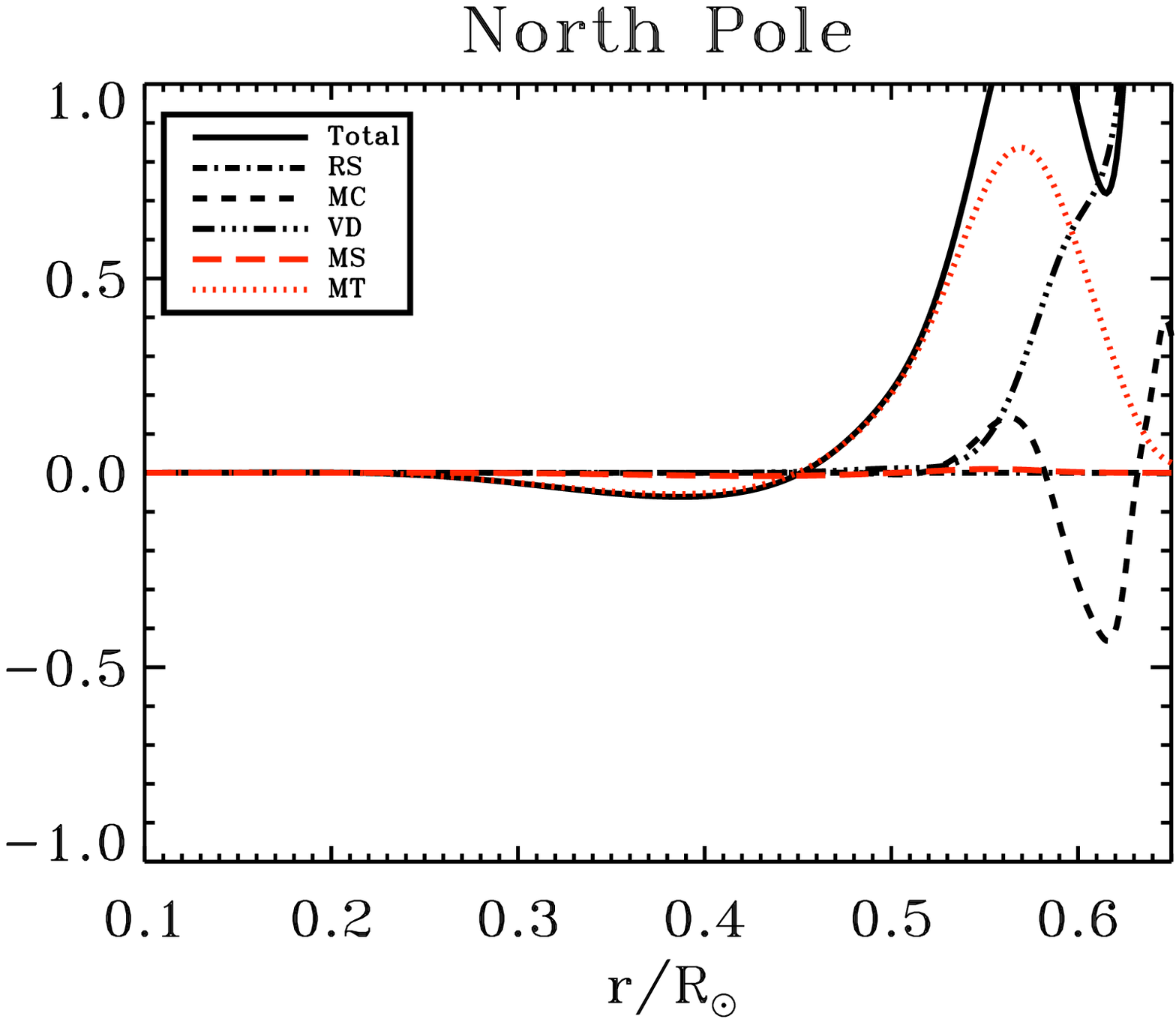}      
  \includegraphics[width=.8\linewidth,angle=90]{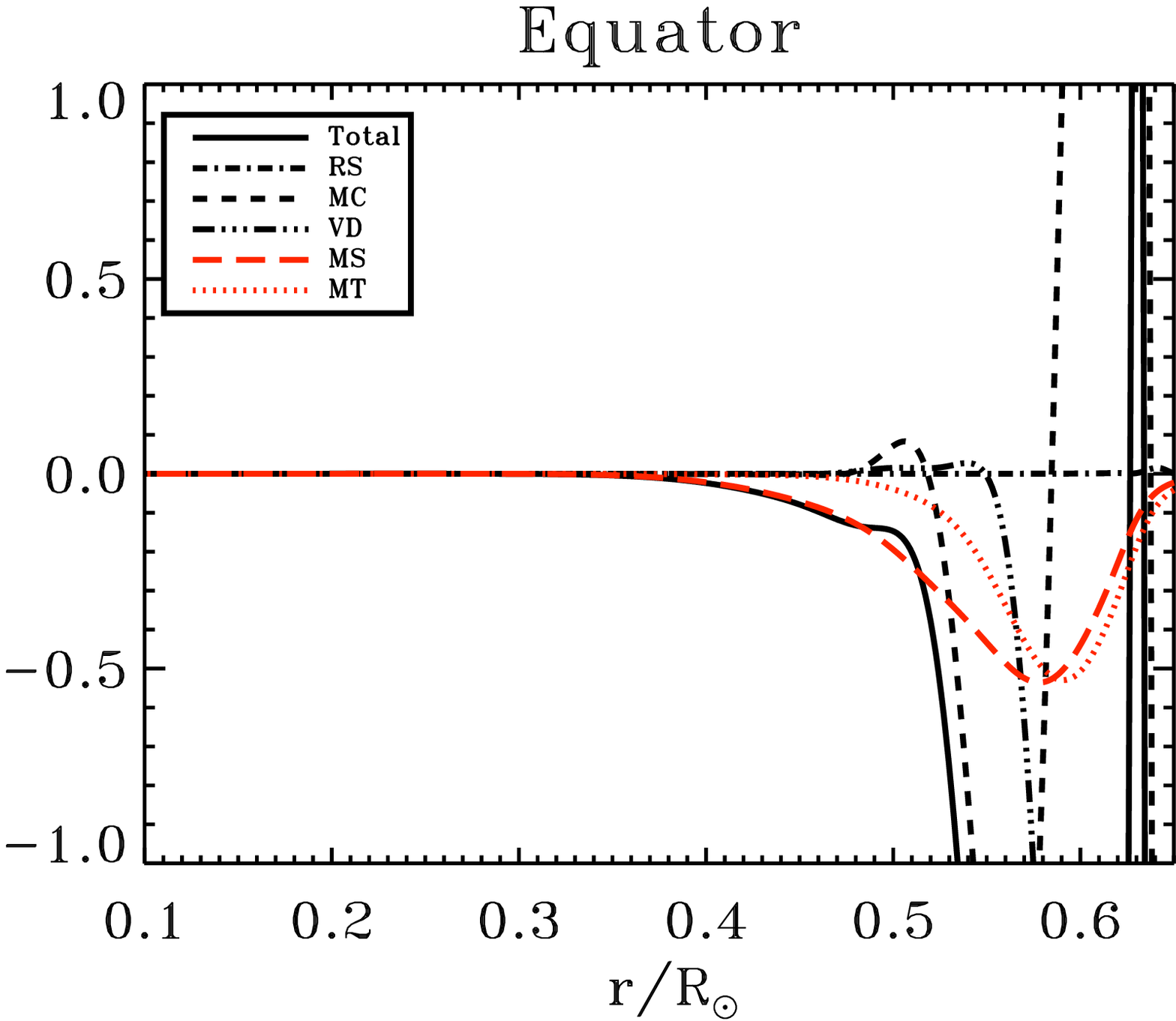}      
 \caption{Time-averaged angular momentum balance ($\mathcal{I}_r$) at the north pole
  $(\theta_1=60,\theta_2=90$, see eq. \eqref{eq:poloidal_flux_ang_mom}) and at the equator
  $(\theta_1=-15,\theta_2=15)$. $\mathcal{I}_r$ is normalized to $10^{12}$.}
  \label{fig:amom_eq_np}
\end{figure}

We plot in Fig. \ref{fig:amom_eq_np} the time-averaged angular momentum balance
near the north pole and near the equator about $75$ convective turnover times after the
introduction of the magnetic field. The magnetic contributions are
highlighted in red and are separated in axisymmetric (Maxwell torque
MT) and non axisymmetric (Maxwell stress MS). At the beginning, those
contributions are exactly zero. Since the field is
initially buried deep in the radiation zone, we first notice that angular momentum
is transported at the top of the radiative zone by (axisymmetric)
Maxwell torque at the north pole (as already noticed in SBZ11). This
again leads to a differentially rotating radiative zone. What is remarkable here
is that we also observe magnetic transport of angular momentum at the
equator. The axisymmetric and non-axisymmetric (torque and stress) components
contribute equally to the inward transport of angular momentum, thus
accelerating the upper radiation zone at low latitude, while Maxwell
torque slows down the high latitude. We can thus conclude that an
oblique field is acting very similarly to a purely aligned dipole.

\section{Conclusions and Perspectives}
\label{sec:consl-persp}

In a previous paper (SBZ11), we tested the \citet{Gough:1998ik} magnetic
confinement scenario of the solar tachocline with a 3D non-linear MHD
model of $90\%$ of the Sun which couples together the convective and
radiative zones. We showed that a fossil axisymmetric magnetic field
does not remain confined in the radiation zone, and thus cannot
prevent the spread of the tachocline.

We showed that the motions (both meridional circulation and
convection) were not strong enough to confine the field and that
upward motions at the base of the convection zone were on the contrary
helping the field to expand inside the convection zone.
Although our numerical experiment is far from representing the
real Sun, we are confident that our parameter regime allows us to
model the main ingredients of the tachocline confinement
scenario. More precisely, we demonstrated in SBZ11 that even if we
consider high magnetic diffusivity, our magnetic Reynolds number is higher than
$1$ in the overshooting/penetration region and in the convection zone.
The magnetic field evolution is thus dominated by advection and not by
diffusion. The overshooting depth of the convective
plumes is likely to be overestimated because our Peclet number is
lower than the solar one, and the penetration of the
meridional circulation may be underestimated
\citep[see][]{Garaud:2008fe,Wood:2011id} in our model. However,
penetration and overshooting are clearly observed
in our model. If more 'realistic' values of the parameters could have
consequences on the polar dynamics, we stress here that it may not
have much influence at the equator, where the confinement of the
magnetic field fails.

The magnetic topology considered in SBZ11 was very simple
(axisymmetric dipole), and the robustness of the results is not
obvious when considering the geometry of the field. We demonstrated in
Sect. \ref{sec:opening-field-at} that an axisymmetric dipole
preferentially diffuses at the location where its maximum radial gradient is,
\textit{i.e.,} at the equator. This has some importance since SBZ11
showed that the lack of confinement specifically at the equator was
responsible for the failure of the \citet{Gough:1998ik} scenario. In
order to test other topologies, we reported in Sect. \ref{sec:diff-magn-topol} the study of an oblique
dipole buried in the radiation zone. A tilted dipole can be
decomposed into an axisymmetric and a non-axisymmetric part, and the
axisymmetric component was shown to evolve similarly to the purely
axisymmetric dipole of SBZ11. As a result, the magnetic confinement
also fails for the tilted dipole.

The non-axisymmetric component of the titled dipole exhibits
interesting three dimensional dynamics and transports angular momentum
at the equator. In order to get rid of the SBZ11 dynamics and to isolate
the effect of the non-axisymmetric dipole, simulations with a purely perpendicular
dipole (\textit{i.e.}, with no axisymmetric component, $\beta=90$ in
Eq. \eqref{eq:dipole_nax_ASH}) will be reported in a future paper.

\begin{acknowledgements}
The authors acknowledge funding
by the European Research Council through ERC grant STARS2 207430
(\url{www.stars2.eu}).
3D renderings in Fig. \ref{fig:3D_ob_dip} were made with SDvision \citep[see][]{Pomarede:2010wp}.
\end{acknowledgements}


\bibliographystyle{astroads}
\bibliography{bib_astro}




\end{document}